\documentstyle [12pt] {article}

\newcommand{\gtwid}{\mathrel{\raise.3ex\hbox{$>$\kern-.75em\lower1ex
\hbox{$\sim$}}}}
\newcommand{\ltwid}{\mathrel{\raise.3ex\hbox{$<$\kern-.75em\lower1ex
\hbox{$\sim$}}}}
\newcommand{\beq}{\begin{equation}}
\newcommand{\eeq}{\end{equation}}
\newcommand{\be}{\begin{equation}}
\newcommand{\ee}{\end{equation}}
\newcommand{\beqs}{\begin{eqnarray}}
\newcommand{\eeqs}{\end{eqnarray}}
\def\({\left (}
\def\){\right )}
\def\AA{\bar{q}_Q}
\def\BB{\bar{q}_{u^c}}
\def\CC{\bar{q}_{d^c}}
\def\DD{\bar{q}_L}
\def\EE{\bar{q}_{e^c}}
\def\FF{q_{H_1}}
\def\GG{q_{H_2}}

\begin{document}

\begin{titlepage}

\begin{flushright}
\begin{tabular}{l}
ITP-SB-95-22    \\
June, 1995
\end{tabular}
\end{flushright}

\vspace{8mm}
\begin{center}
{\Large \bf U(1)$_A$ Models of Fermion Masses Without a $\mu$ Problem}

\vspace{4mm}
\vspace{16mm}

Vidyut Jain \footnote{email: vid@max.physics.sunysb.edu}
and Robert Shrock\footnote{email: shrock@max.physics.sunysb.edu}

\vspace{6mm}
Institute for Theoretical Physics  \\
State University of New York       \\
Stony Brook, N. Y. 11794-3840  \\

\vspace{20mm}

{\bf Abstract}
\end{center}

We discuss the connection between models of fermion masses and mixing
involving a string-motivated flavor/generation U(1)$_A$ gauge symmetry and
the $\mu$ term.  We point out that in a certain class of such models the
flavor physics can provide an appealing solution to the $\mu$ problem,
naturally yielding a $\mu \sim O(m_{_W})$.  A simple relationship between the
$U(1)_A$ charge $q_{H}$ of
the $\mu$-term and the average generational $U(1)_A$
charges of the down quark and leptonic sectors is derived. Finally, we
construct an explicit model illustrating our results.

\vspace{35mm}

 \end{titlepage}
\newpage
\setcounter{page}{1}
\pagestyle{plain}
\pagenumbering{arabic}
\renewcommand{\thefootnote}{\arabic{footnote}}
\setcounter{footnote}{0}

   Two of the most outstanding challenges at present in particle physics are
to understand (a) the observed pattern of fermion masses and mixing; and (b)
the hierarchy between the electroweak and Planck mass scales.  The standard
model (SM) can accomodate, but not explain these fermion masses and mixing.
Renormalizable supersymmetric extensions of the standard model can
stabilize the Higgs sector against large radiative corrections, but this, by
itself, is not enough to explain property (b) because of the
$\mu$--problem.  This problem concerns the fact that
the term $\mu H_1 H_2$ in the superpotential $W$, where
$H_i$, $i=1,2$ are the $Y=-1,1$ Higgs chiral superfields in the minimal
supersymmetric standard model (MSSM), is supersymmetric, and hence there
is no apparent reason why its coefficient,
$\mu$, should have anything to do with the scale of supersymmetry breaking
(which, in realistic theories is of order the electroweak scale,
$v_{_{EW}} \simeq 250$ GeV).  If, as is allowed by supersymmetry itself,
$|\mu| >> v_{_{EW}}$, this would destroy the hierarchy, and, indeed, for
a phenomenologically acceptable model, $|\mu|$ must be of order $v_{_{EW}}$.
On the other hand, one cannot forbid the $\mu$ term
completely, for this would lead to a phenomenologically unacceptable
massless goldstone boson.\footnote{For a recent
review of the $\mu$ problem, see Ref. \cite{murev}.}

   In this paper we discuss an appealing class of models yielding a
unified explanation of both of these outstanding problems (a) and (b). To
introduce these, we first note that recently there has been strong interest
in models of fermion masses
and mixing based on a string-motivated, flavor- and generation-dependent
U(1)$_A$ gauge symmetry \cite{ibanez}--\cite{dps} in
the context of a supergravity theory reducing at $E <<
\bar M_P$ to the MSSM.\footnote{Here,
$\bar M_P \equiv (8 \pi G_N)^{-1/2} = 2.44 \times 10^{18}$ GeV is the (reduced)
Planck mass.}  The U(1)$_A$ symmetry has field-theoretic anomalies
which are cancelled by a Green-Schwarz mechanism \cite{gs}; as a result, this
symmetry is spontaneously broken, at a calculable high mass scale
somewhat below the string scale \cite{dsw}.

   When one embeds global supersymmetry in a supergravity theory, the
$\mu$--parameter has an additional contribution from the K\"ahler potential,
$K$ \cite{gm}
\be
\tilde{\mu}(M,M^{\dag}) H_1 H_2
\label{muink}
\ee
where the parenthesis indicates the dependence of $\tilde \mu$ on generic
fields $M$ which aquire large vacuum expectation value (vev's). Upon
supersymmetry
breaking, the second term can lead to a contribution to an effective
superpotential $\mu$-term of the form
\be
\tilde{\mu}(<M>,<M^{\dag}>) m_{3/2} H_1 H_2
\label{mueffinw}
\ee
in $W_{eff}$ .
Here $m_{3/2}$ is the gravitino mass, which is related to the scale of
supersymmetry breaking,
\be
\mu \sim v_{_{EW}} \sim m_{3/2}
\label{musimv}
\eeq
If the superpotential contribution to the $\mu$--term is zero, it becomes
possible to generate a $\mu$ term of an acceptable size (\ref{musimv})
in $W_{eff}$, via eq. (\ref{mueffinw}), from K\"ahler couplings.
In many string-inspired effective supergravity
models the superpotential contribution to the $\mu$-term is absent at
tree level because of modular invariance, but can be generated by
nonperturbative effects in the hidden sector
\cite{murev,muwork,anton}. Unfortunately, it is difficult to ascertain whether
this nonperturbative contribution is of an acceptable size \cite{anton}.
Moreover, in generic supergravity models there are even more serious problems
with the $\mu$ term, and hence stabilization of the hierarchy, than in
renormalizable supersymmetric models.  This is because of nonrenormalizable
couplings in $K$ which can lead to quadratically divergent loop corrections
to $\mu$ \cite{vj,bagger} (see also \cite{gj}).

However, as was first pointed out in Ref. \cite{vj}\footnote{See
discussion after eq. (24) in Ref. \cite{vj}.}, in models with a $U(1)_A$ gauge
symmetry, such a destabilization
can be avoided and furthermore, the natural value of the $\mu$-parameter
is determined by the couplings allowed by the gauge group, so
that in such models one can achieve a solution of the
$\mu$ problem.
Moreover, as we discuss here, this solution connects the
physics of fermion masses and mixing with that of the hierarchy in a
fundamental way. This idea was embodied in new solutions of the
anomaly cancellation conditions discovered in Ref. \cite{js}
which had the property that the sum of the U(1)$_A$ charges of the Higgs,
$q_{_H} \equiv q_{_{H_1}}+q_{_{H_2}} \ne 0$.    If and only if
$q_{_H} \ne 0$, the $\mu$--term cannot, by itself, appear in either
$W$ or $K$. However, it may appear in combination with powers of various other
chiral superfields such that the total $U(1)_A$ charge
is zero, and in this case the natural order of the effective low energy
$\mu$--parameter,
which is calculable in terms of the $U(1)_A$ charges of these
various fields, can be naturally of order the electroweak scale.

   We first briefly review U(1)$_A$ models of fermion masses and mixing.
One of the most puzzling aspects of the known fermion masses is that
if one assumes, as in the SM, that they arise from
conventional, dimension-4 Yukawa operators, then
the associated Yukawa couplings for all of these fermions
except the top quark are all much smaller than a typical small coupling like
$e=\sqrt{4\pi\alpha} \simeq 0.3$, without any explanation.  In these
models, this feature of the first two generations is explained in an elegant
way, since they arise from higher-dimension, nonrenormalizable
operators.\footnote{An early paper noting the importance of such operators
for fermion mass relations is Ref. \cite{eg}.}
This is appealing, because these operators are generically present in the
supergravity theory which forms the field-theory limit of the presumed
underlying string theory for energies $E << M_{str}$ (where
$M_{str} = 2(\alpha')^{-1/2} = g\bar M_{P}$ denotes the string scale)
with $c$-number coefficients proportional to the requisite
inverse powers of $M_{str}$. Via
vacuum expectation values of the scalar components of certain chiral
superfields, which we shall denote generically as $v$, these
higher-dimension operators can yield contributions to effective
dimension-4 Yukawa interactions which are suppressed by powers of the ratio
$\epsilon \sim v/M_{str}$.  An important aspect of this approach is that
$\epsilon$ is calculable; the U(1)$_A$ symmetry-breaking scale $v$ is
given by \cite{dsw} $v^2 \simeq (M_{str})^2/(192 \pi^2)$, whence
$\epsilon \sim \lambda^2$ (where $\lambda=|V_{us}| \approx 0.22$), so that
$\epsilon$ is in the right range to explain the fermion mass
hierarchies\footnote{
Here the running masses are used, normalized at the same scale.  Note that
$m_d m_s m_b = \lambda^6 m_b^3$, $m_e m_\mu m_\tau = \lambda^8 m_\tau^3$, and,
since $m_b(m_b) \sim \lambda^{-2/3} m_b(v)$ with $m_b(v) \simeq m_\tau(v)$,
there follows the well-known relation
$\det M_d(v) \simeq \det M_L(v)$ which we shall use below.}
\beq
 m_u/m_c \sim \lambda^5, \quad m_c/m_t \sim \lambda^4, \quad
m_d/m_s \sim \lambda^2, \quad m_s/m_b \sim \lambda^2
\label{hier1}
\eeq
\beq
 m_e/m_\mu \sim \lambda^4, \quad m_\mu/m_\tau \sim \lambda^2
\label{hier2}
\eeq
The U(1)$_A$ symmetry also forbids certain chiral
superfield cubic couplings and hence produces zeros in some entries of the
Yukawa matrices for the resultant dimension-4 Yukawa terms.  This, in turn,
enables one to explain why the off-diagonal elements of the
Cabibbo-Kobayashi-Maskawa (CKM) matrix are small by allowing one to express
them in terms of small quark mass ratios $m_d/m_s$, $m_u/m_c$,
etc.\footnote{The
latter result was also achieved in early models of quark mass
matrices \cite{wf}; however, since these used only dimension-4 Yukawa
couplings, they had to put in the fermion mass hierarchy by hand.}

 A further puzzling property
of the observed fermion mass spectrum is
that, even if one restricts oneself just to the third generation, $m_b$ and
$m_\tau$ are $<< m_t$, again with no explanation.  A general class of
U(1)$_A$ models was discussed in Ref. \cite{js}, and a specific string
construction in Ref. \cite{far}  which account for this by
means of a far-reaching
hypothesis, namely, that even the largest, $j,k=3,3$ elements of the
(effective) down-quark and charged lepton Yukawa matrices $Y^{d}$ and
$Y^{L}$ actually arise from higher-dimension operators.

   We note an additional appeal of the U(1)$_A$ symmetry.
One might, {\it a priori}, try to restrict the form of the fermion Yukawa
matrices using a global symmetry.  However, in addition to the unesthetic
nature of global symmetries as fundamental and problems with
unwanted goldstone bosons resulting from spontaneous symmetry breaking,
there is also the problem that they are, in general, broken by quantum
gravity, even at the semi-classical level \cite{qgbk}.

  There are at least two worthwhile ways to study U(1)$_A$--based models of
low energy physics\footnote{One
should also note that efforts have been made to describe
the origin of this pattern in terms of new physics operative at a much lower
mass scale, e.g., 1-100 TeV, in theories involving dynamical electroweak
symmetry breaking. See also Ref. \cite{nir}}:
(a) one can study specific
scenarios suggested by particular string constructions, as in
Ref. \cite{far,lykken}; (b) alternatively, one can avoid committing
oneself to any particular string constructions, and instead study general
solutions to the consistency conditions implied by the Green-Schwarz
mechanism.  We continue to follow the second approach
here, since it allows one to explore in a very general way what can and cannot
be achieved with such models.
As in Ref. \cite{js}, we shall study the simplest extension of the SM which
may be able to account for the observed fermion masses and the $\mu$-term,
i.e. one in which  the observable sector gauge group is just
SU(3) $\times$ SU(2) $\times$ U(1)$_Y \times$ U(1)$_A$ and
the matter content includes just the usual SM fields together with some
additional SM--singlet fields which can carry $U(1)_A$ charge.
The model we study is a minimal supergravity extension
of the SM with enough new physics built in to have the potential to explain
low energy data.
Our objective is to explore the implications of such a minimal
model.  Indeed, we find that such a model can be phenomenologically
viable.\footnote{
In generic string-motivated models there are extra (nonanomalous) $U(1)$
gauge symmetries, as well as discrete symmetries and nontrivial kinetic
normalizations for the matter fields.}

   As part of our analysis, we will derive a simple linear relation
between $q_{_H}$ and the average (generational) $U(1)_A$ charges of
the SM down-quark and lepton sectors.
This result holds independent of (i) whether the $U(1)_A$ charges
of the superpotential Yukawa terms are assumed to be symmetric in
generation-space or not and (ii) the spectrum of SM--singlet $U(1)_A$ charged
chiral superfields, which will be denoted generically as $\chi, \
\tilde \chi$, etc.  We consider the general
case where the effective Yukawa matrices in the up, down, and charged
lepton sectors are not necessarily symmetric in generation indices.
We find that a K\"ahler potential origin for the effective $\mu$-parameter
may well be more
natural than a superpotential origin. Indeed, if only one SM--singlet
field $\chi$ gets a vev due to $U(1)_A$ breaking, then an
acceptable $\mu_{eff}$ must come from $K$ (unless $q_{_H}=0$)
whereas if just two SM--singlet fields $\chi,\tilde{\chi}$ with opposite
$U(1)_A$ charges get vev's, then the largest contribution to $\mu_{eff}$
is always from $W$ and an acceptable value is generated only if
$|q_{_H}|\sim 11 |q(\chi)|$. When all SM--singlet fields $\chi$ which
get vev's carry the same sign $U(1)_A$ charge then an acceptable
$\mu_{eff}$ may originate from either $K$ or $W$, but it appears  more easily
 from $K$.

 If the Yukawa matrix charge assignments are symmetric in
generation space, the number of free parameters reduces greatly and stronger
statements are possible. In fact, in this case, it is difficult to reproduce
observed fermion mass hierarchies if (i) both top and bottom masses arise
 from renormalizable superpotential couplings, (ii) effective Yukawa couplings
are due to the vev of just one SM--neutral $U(1)_A$ charged field or
(iii) effective Yukawa couplings are due to the vev's of just two
SM--neutral fields with opposite (nonzero) $U(1)_A$ charges. It is possible to
avoid problems associated with these three cases if the bottom mass arises
 from a nonrenormalizable superpotential coupling and effective Yukawa
couplings are due to several SM--singlet fields which all carry the same
sign (nonzero) $U(1)_A$ charge. When there are just two such fields we
show that an acceptable $\mu$-parameter cannot be achieved from
$W$ if $\tan\beta$ is not allowed to get either very large or very small.

   To begin, we assign arbitrary $U(1)_A$ charges to all fields. This
introduces 17 parameters for the 3 generation minimal standard model, plus
additional parameters for SM singlets \cite{js}. We use the parametrization
\beq
q(Q_i)=\bar q_Q + \alpha_i \ , q(u^c_i) = \bar q_{_{u^c_i}}+\beta_i \ ,
\ q(d^c_i)=\bar q_{_{d^c}}+\gamma_i \ , \ q(L_i)=\bar q_L + a_i \ , \
q(e^c_i)=\bar q_{e^c}+b_i
\label{qs}
\eeq
where $\bar q(f) = (1/3)\sum_{i=1}^3 q(f_i)$ denotes the generational average
U(1)$_A$ charge of $f$, and thus
\beq
\sum_i \alpha_i = \sum_i \beta_i = \sum_i \gamma_i = \sum_i a_i = \sum b_i = 0
\label{alphaetc}
\eeq

  With the standard normalization of $U(1)_Y$ hypercharge, $Y(Q)=1/3$,
$Y(u)=-4/3$, $Y(d)=2/3$, $Y(L)=-1$, $Y(e)=2$, $Y(H_2)=1$, and $Y(H_1)=-1$, the
mixed anomaly coefficients are
\beq
c_1={\rm Tr}(T_a T_Y^2)={\AA\over 2}+4\BB+\CC+{3\over 2}\DD+3\EE+{\FF\over 2}
     +{\GG\over 2}
\label{c1anom}
\eeq
\beq
c_2={\rm Tr}(T_a T_{SU(2)}^2)= {9\over 2}\AA+{3\over 2}\DD+{\FF\over 2}
     +{\GG\over 2}
\label{c2anom}
\eeq
and
\beq
c_3={\rm Tr}(T_a T_{SU(3)}^2)=3\AA+{3\over 2}\BB +{3\over 2}\CC
\label{c3anom}
\eeq
Since these anomaly coefficients are linear in the U(1)$_A$ charges and involve
sums over all generations, they depend only on average U(1)$_A$ charges, as
indicated.  An immediate consequence is that
\beq
 c_1+c_2-{8\over 3} c_3=  q_{_H} - 3(\AA+\CC-\DD-\EE)
\label{chrel}
\eeq
The anomaly cancellation by the Green-Schwarz mechanism requires \cite{ibanez}
$c_i/c_j=k_i/k_j$ where the $k_i$ are the levels for the Kac-Moody algebra on
the string worldsheet which determine the gauge couplings for each of the
factor groups U(1)$_Y$, SU(2), and SU(3) by
$g_i^{-2}=k_i \langle Re(s)\rangle$ (at $M_{str}$), where, in turn, $s$ is the
dilaton/axion.  The unification of gauge couplings in the MSSM \footnote{
For a recent discussion of the current status of MSSM gauge
unification, see Ref. \cite{kane}.}
requires
$g_1^{-2}=(5/3)g_i^{-2}$, $i=2,3$ and hence (with $k_2=k_3=1$ to avoid
exotics),
\beq
 c_1 : c_2 : c_3 = \frac{5}{3} : 1 : 1
\label{crel}
\eeq
 From this it follows that
\beq
  c_1+c_2-{8\over 3} c_3=0, \ i.e.,
\quad  q_{_H}=3(\AA+\CC-\DD-\EE)
\label{hchargerel}
\eeq
where, as before, $q_{_H}=q_{_{H_1}}+q_{_{H_2}}$.  Thus, the U(1)$_A$
charge $q_{_H}$ of the $\mu$ term $H_1H_2$ is determined in terms of the
generational-average charges of the matter fermions.
Later we will give another expression for the RHS which will
have important consequences for the effective $\mu$-parameter.

A further requirement of anomaly cancellation is that
\beq
0={\rm Tr}(T_a^2 T_Y)=3\AA^2-6\BB^2+3\CC^2-3\DD^2+3\EE^2+\GG^2
        -\FF^2 +\Delta
\label{anom0}
\eeq
where $\Delta$ is quadratic in parameters $\alpha_i,\beta_i,\gamma_i, a_i,b_i$
which do not depend on average (barred) charges. In general, there is
no simple solution to eq. (\ref{anom0}); however, when $\Delta=0$,
eq. (\ref{anom0}) imposes a simple quadratic constraint on the average
$U(1)_A$ charges.  $\Delta$ vanishes identically when additional symmetry
requirements are imposed, e.g. when the Yukawa charge assignments are
required to be symmetric in flavor space, as in \cite{ir,js} so that
$\alpha_i=\beta_i=\gamma_i, a_i=b_i$.  We recall the general solutions to the
full set of anomaly constraints found in Ref. \cite{js} for the
$\Delta=0$ case:
\beq
\begin{array}{ccccccc}
 \AA & \BB & \CC & \DD & \EE & \GG & \FF \\
  x & x & y & y & x & z & -z \\
  x & x & {y\over 2}-{z\over 2} & y & x & -{3\over 2}y -{1\over 2} z &
   -z \\
  \; x+v \; & \; x+2v \; & \; y+w \; & \; y \; & \; x \; &
   \; 3v+3w+z \; & \; -z \;
\end{array}
\label{sols}
\eeq
The first two solutions  describe two distinct 3--parameter
family of solutions to the constraints. The last solution, with
$v\neq 0$, describes a 4 parameter family of solutions with $x$ given by
\beq
 0= -2v^2+2w^2+3v(w-x)+vz+w(y+z)
\label{constraint}
\eeq
The first solution in (\ref{sols}) was already given in \cite{ir}, while the
latter two were new in Ref. \cite{js}.  These two new solutions which we
discovered, and which
allow $q_{_H}=q_{H_1}+q_{H_2}\neq 0$, can play an important
role in constraining the $\mu$-parameter.

  Labelling the U(1)$_A$ charges of the cubic superfield terms as
\beq
q(Q_3 u^c_3 H_2)=\delta_t \ , \quad q(Q_3 d^c_3 H_1)=\delta_b \ ,
 \quad q(L_3 e^c_3 H_1)=\delta_\tau
\label{qcubic}
\eeq
we have
\begin{eqnarray}  \AA+\BB+\GG+\alpha_3+\beta_3 &=& \delta_t, \nonumber \\
    \AA+\CC+\FF+\alpha_3+\gamma_3 &=& \delta_b, \nonumber \\
    \DD+\EE+\FF+a_3+b_3 & = & \delta_\tau
\label{qqrel}
\end{eqnarray}
The U(1)$_A$ charges Yukawa coupling charge matrices are given
by
\beq
q(Q_i u^c_j H_2)=\delta_t + \alpha_i-\alpha_3+\beta_i-\beta_3
\label{quch2}
\eeq
\beq
q(Q_i d^c_j H_1)=\delta_b + \alpha_i-\alpha_3+\gamma_i-\gamma_3
\label{qdch1}
\eeq
\beq
q(L_i e^c_j H_1)=\delta_\tau + a_i-a_3+b_i-b_3.
\label{qech1}
\eeq
For the symmetric charge assignment case, $\beta_i=\gamma_i=\alpha_i,b_i=a_i$,
the charge matrix structures (\ref{quch2})-(\ref{qech1}) imply
\beq
 q(Q_i d^c_j H_1) = q(Q_i u^c_j H_2)+(\delta_b-\delta_\tau).
\label{quqdrel}
\ee
 From this we deduce:

\begin{enumerate}

\item
 If $\delta_t=\delta_b$ then the up-quark $U(1)_A$ charge matrix is identical
to the down-quark $U(1)_A$ charge matrix. Hence, we have $Y^u_{ij}
\sim Y^d_{ij}$ which cannot fit experimental data.  This includes the case
where the top and bottom masses both arise from renormalizable couplings.

\item
  If superpotential couplings of the quark sector to a single SM--singlet
field which receives a vev are responsible for all effective (low-energy)
Yukawa matrices, then we again must have
$Y^u_{ij}\propto Y^d_{ij}$ which is in contradiction with data.

\item
  If superpotential couplings of the quark sector to two SM--singlet
fields which have opposite $U(1)_A$ charges and receive vev's
 are responsible for all (effective) Yukawa matrices, then we
also must have $Y^u_{ij}\propto Y^d_{ij}$, again in disagreement with data.

\end{enumerate}

Therefore if one wishes to explain low energy Yukawa data via the U(1)$_A$
symmetry, and one wants either $\delta_t=\delta_b$ $or$ invokes SM fermion
couplings to only a single SM--singlet field or two SM--singlet fields with
opposite $U(1)_A$ charges, then one must relax the condition of symmetric
charge assignments (as in \cite{br,dps}).

   However, there is no {\it ab initio} reason to use just a single
SM--singlet field (or a pair of oppositely charged SM--singlets),
and, indeed, as
we previously demonstrated \cite{js}, in the context of symmetric
textures, the above restrictions can be avoided if one considers, for example,
two SM--singlet, U(1)$_A$-charged fields $\chi,\chi'$ which do not have
opposite charges. We will exhibit a new model of this type
which is able to fit all low energy mass data.

   Let us next define $\Sigma$ as
the  diagonal sum of $U(1)_A$ charges in a Yukawa charge
matrix. Hence,
\begin{eqnarray} \Sigma_u &=& 3(\delta_t-\alpha_3-\beta_3),
\nonumber \\
                 \Sigma_d &=& 3(\delta_b-\alpha_3-\gamma_3),
\nonumber \\
   \Sigma_L &=& 3(\delta_\tau-a_3-b_3).
\label{sigmavalues}
 \end{eqnarray}
Note that $\Sigma$ is also the $U(1)_A$ charge of any term in the determinant
of a Yukawa coupling matrix, e.g. $\Sigma_d=q(\det[Q_i d_j^c H_1])$, and
similarly for $\Sigma_L$ and $\Sigma_u$.  From eqs. (\ref{chrel}) and
(\ref{qqrel}), we get
\beq
   c_1+c_2-{8\over 3}c_3 = q_{_H}+\Sigma_L-\Sigma_d
\label{csigma}
\eeq
or, for the condition (\ref{crel}),
\beq
   q_{_H}=\Sigma_d-\Sigma_L.
\label{qhsigmarel}
\eeq
As we will see, this form of eq. (\ref{hchargerel}) makes the analysis of
the relation between the effective $\mu$-parameter and fermion mass
matrices quite transparent in many cases of interest.

It is interesting to note that there is a general relation between
$k_1/k_3$ and the Yukawa and $\mu$-term
superpotential charges. From eqs. (\ref{c3anom}), (\ref{qqrel}),
and (\ref{sigmavalues}), we have
\beq 2c_3= \Sigma_u+\Sigma_d-3q_{_H}. \ee
For $k_2=k_3$, hence $c_2=c_3$, we have
\beq
   c_1-{5\over 3}c_3= q_{_H}+\Sigma_L-\Sigma_d.
\label{anothercsig}
\eeq
Together, these last two equations yield
\beq {k_1\over k_3} = {c_1\over c_3} =
   {6\Sigma_L-\Sigma_d + 5\Sigma_u - 9 q_{_H} \over
     3\Sigma_d +3\Sigma_u - 9q_{_H} }
\label{kratio}
\eeq

   In models involving a single SM--singlet
U(1)$_A$--charged field, one can derive the relation $\sin^2\theta_w=3/8$
if one has $q_{_H}=0$ and $c_2=c_3$ \cite{br}. (This can be seen from
our eq. (\ref{kratio}); the condition $\det Y^L \sim \det Y^d$ implies
$\Sigma_L \sim \Sigma_d$, so the RHS of eq. (\ref{kratio}) $=5/3$ for
$q_{_H}=0$.) However, the property that
$q_{_H}=0$ in such models means that they do not solve the $\mu$ problem.
In contrast, the relations we have derived, (\ref{hchargerel}) and
(\ref{qhsigmarel}), are independent
of the number of $U(1)_A$ charged fields and can be used to constrain
$U(1)_A$ charges in more general situations. Our approach is to assume
canonical Kac-Moody levels $k_2=k_3=1$ and $k_1=5/3$ for which
$c_1+c_2-{8\over 3}c_3=0$ and use eqs. (\ref{hchargerel}) and
(\ref{qhsigmarel}) to analyze relations between $q_{_H}$ and fermion masses.
The $\mu$-term charge $q_{_H}$
is only zero if $\Sigma_d=\Sigma_L$, which is sufficient,
but not necessary, to satisfy data on fermion masses and mixing.  The point
here is that when one says the $i,j$
entry in an effective Yukawa matrix is of order $\lambda^n$, this means
that the dimensionless coefficient $c_{ij}$ multiplying
$\lambda^n$ lies within the range $\lambda \le c_{ij} \le \lambda^{-1}$.  This
finite range in each of the coefficients produces an intrinsic uncertainty in
the actual, as opposed to formal, power of $\lambda$ describing the determinant
of the Yukawa matrix.  Indeed, because of this uncertainty, the fermion mass
data can be satisfied without requiring that $\Sigma_L=\Sigma_d$ as long as the
ratio between these determinants is within the range induced by the above
dimensionless coefficients.  Hence, even in such a model,
a small $\mu$-parameter may be obtained by requiring $q_{_H}$ to be small and
nonzero so that only a K\"ahler potential contribution to $\mu_{eff}$ is
obtained.

   If a model has two SM--singlet fields $\chi,\tilde{\chi}$
with opposite $U(1)_A$ charges $x,-x$, then one needs
$|q_{_H}|\sim 11 |x|$ to get an acceptable value for $\mu_{eff}$.
This is because
if $q_{_H}$ is not an integer multiple of $x$ then no $\mu$-term is allowed,
and if $q_{_H}$ is an integer multiple of $x$ then the largest contribution
is always from $W$ and is either of the form
$\chi^{|q_{_H}|/|x|} H_1 H_2$ or
$\tilde{\chi}^{|q_{_H}|/|x|}H_1 H_2$.
Assuming vev's of $O(\lambda^2) \bar M_P$, we then
find $\mu\sim \lambda^{2|q_{_H}|/|x|} \bar M_P$.

   We now study the case of two fields $\chi,\chi'$ with same--sign
$U(1)_A$ charges. With no loss of generality, we take
\be
q[\chi]=1,\quad q[\chi']=\alpha >1.
\label{qchoices}
\ee
We have three possibilities:

\begin{enumerate}

\item

$q_{_H}=0$. If one were to choose this value, the model would have an
unresolved $\mu$--problem, so we avoid this choice.

\item

 $q_{_H}<0$. If $q_{_H}=-n-m\alpha$ for nonnegative integers $n,m$, then $W$
contains the term
\be
{\chi^n (\chi')^m \over (\bar M_P)^{(n+m-1)} } H_1 H_2 ,
\label{wtermhm}
\ee
so that upon $U(1)_A$ breaking,
\be
\mu\sim \lambda^{2(n+m)} \bar M_P.
\label{wtermhm2}
\ee
In this case, unless $n+m$ is sufficiently large, one naturally has
too large a $\mu$--term. On the other hand, if $q_{_H}<0$ but $q_{_H}\ne
-n-m\alpha$ for nonnegative integers $n,m$, then the superpotential
contribution to the total $\mu$-term is zero.

\item
  $q_{_H}>0$. In this case there can be no superpotential $\mu$--term; however
the K\"ahler potential contains the terms
\begin{eqnarray}
 (a) & & (\chi^{\dag})^n(\chi'^{\dag})^m H_1 H_2
     \;\;{\rm if}\;\; q_{_H}=n+m\alpha, \nonumber \\
  {\rm or}\;\; (b) & & \chi(\chi'^{\dag})^m H_1 H_2 \;\;{\rm if}\;\;
      q_{_H}=-1+m\alpha, m\geq 1, \nonumber \\
     {\rm or}\;\; (c) & & (\chi^{\dag})^n\chi' H_1 H_2
      \;\; {\rm if}\;\; q_{_H}=n-\alpha, n\geq 2,
\label{nextchoices}
\end{eqnarray}
for some nonnegative integers $n,m$. Upon SUSY and $U(1)_A$
breaking, these give contributions to the effective $\mu$--term
\begin{eqnarray}
   (a) & & \mu_{eff}\sim m_{3/2}\lambda^{2(n+m)}, \nonumber \\
   (b) & & \mu_{eff}\sim m_{3/2}\lambda^{2m+2}, \nonumber \\
   (c) & & \mu_{eff}\sim m_{3/2}\lambda^{2n+2},
\label{mucontributions}
\end{eqnarray}

\end{enumerate}
If one of the conditions $(a)$ through $(c)$ above
is not satisfied, then there is
no K\"ahler potential contribution to $\mu$. Otherwise, if the
only contribution to the effective $\mu$-term is from $K$, it will will be
too small unless $n$ and $m$ are sufficiently small. In fact,
$\mu\sim v_{_{EW}}$ suggests that only the cases
\be q_{_H}=1,\quad {\rm or}\quad q_{_H}=\alpha\ee
(corresponding to $K$ containing $\chi^{\dag}H_1 H_2$ or
$\chi'^{\dag} H_1 H_2$,
respectively), which both give $\mu_{eff}\sim m_{3/2}\lambda^2$, can naturally
yield an acceptable value of the $\mu$-parameter.

It appears considerably more difficult to construct models where an
acceptable  $\mu$--term arises from the superpotential. If we require
\be
\det Y^d \sim \det Y^L \sim \lambda^{2d}
\label{ydyl}
 ,\ee
we can derive
\be |q_{_H}|  \leq (\alpha-1)d,
\label{adhrel}
\ee
by considering upper and lower bounds on the determinants,
\be \lambda^{-2\Sigma_d} \le \det Y^d \le \lambda^{-2\Sigma_d/\alpha},\quad
    \lambda^{-2\Sigma_L} \le \det Y^L \le \lambda^{-2\Sigma_L/\alpha},
\label{determinants}
 \ee
and using (\ref{qhsigmarel}).
It follows that if there is a nonzero superpotential $\mu$-term then
$q_{_H}<0$ and
\be \mu  \geq  O(\lambda^{2(\alpha-1) d}) \bar M_P.
\label{mump}
\ee
Since one expects $d\sim 7$, an
acceptable superpotential contribution to $\mu_{eff}$ requires
$\alpha \ge 2.5$.

In models with multiple SM--singlet fields with same--sign $U(1)_A$ charges,
this requires the largest magnitude $U(1)_A$ charge to be at least 2.5 times
greater than the smallest magnitude $U(1)_A$ charge. To realize how
restrictive this can be, we first note that (\ref{mump})
assumes the $\mu$-term couples exclusively
to the SM singlet with charge 1 (i.e. the smallest $U(1)_A$ charge). If,
instead, it couples only to the SM singlet with charge $\alpha>1$, then we
can never produce a sufficiently small superpotential $\mu_{eff}$. Thus, in
general, we expect $\alpha$ to be much bigger than 2.5. Furthermore, there can
be restrictions on $\alpha$ from data on fermion masses and mixing. For
example, in the symmetric case with just two additional fields we
can argue that $\mu_{eff}$ should originate from $K$
in order to satisfy fermion mass data and $\tan\beta$ not too large
(or small). To do this, we concentrate on the two most massive
generations. For simplicity we assume the top mass arises from a renormalizable
operator, i.e. $\delta_t=0$.  From (\ref{quch2}) we note
that in order to have $m_c/m_t\sim\lambda^4$, i.e.
$Y^u_{22}Y^u_{33}-Y^u_{23}Y^u_{23}\sim \lambda^4 (Y^u_{33})^2,$, it is
necessary that
$q(H_2 Q_2 u^c_3)=\alpha_1+2\alpha_2=-1,-\alpha$ or $-(1+\alpha)/2$.
If $\delta_b=0$, then the mass $m_b$
also arises from a renormalizable
operator, i.e. the small value of $m_b/m_t$ must be entirely accounted for
by a large $\tan\beta$. If, however, $\delta_b=-1$ or $-\alpha$, then
$Y^d_{33}/ Y^u_{33} \sim \lambda^2$
and one does not need either a very large or
small value of $\tan\beta$. This
leads to six possible charge assignments for the
up and down-quark Yukawa couplings of the the two highest generations in
terms of the unkown parameter $\alpha$. One can then show that the only
solutions with $m_s/m_b\sim\lambda^2$ have $\alpha=2,{3\over 2}, {2\over 3}$
or ${1\over 2}$. Therefore the bound $\alpha\geq 2.5$
cannot be achieved, and hence
an acceptable $\mu_{eff}$ can only arise from $K$; if there is a contribution
 from $W$ it will be too large. (We must require $q_{_H}=1$ or
$q_{_H}=\alpha$ in order to get a sufficiently large $\mu_{eff}$ from $K$.)

    We next construct a model with symmetric Yukawa matrices and
$q_{H_1}+q_{H_2}\neq 0$ which serves as an explicit example of how one can both
solve the $\mu$ problem and account for fermion masses and mixing with the
U(1)$_A$ symmetry.  It incorporates and extends our results in Ref. \cite{js}.
The model assumes symmetric U(1)$_A$ charge assignments for simplicity
(but could be generalized to the case of asymmetric charges).  It provides a
fundamental explanation for why even in the third generation, the masses $m_b$
and $m_\tau$ are much less than the electroweak scale: these masses are
generated by higher-dimension operators, in contrast to the top mass, which
arises from a dimension-4 operator.  Hence,
$Y^d_{33}$ and $Y^L_{33}$ are both $\sim \lambda^2 Y^u_{33}$. We
found a solution with two SM--neutral fields $\chi,\chi'$
with same sign $U(1)_A$ charge.
Normalizing the charge of one field, $\chi$, to be 1, our solution is
\be
q[\chi'] = {3\over 2},\quad \alpha_1=-{4\over 3}, \quad
   \alpha_2={1\over 6}, \quad \delta_t=0, \quad \delta_b=-1
\label{chargechoice}
\ee
which leads to
\be
Y^u\sim\left ( \begin{array}{ccc}
      \lambda^8 & \lambda^6 & \lambda^4 \\
      \lambda^6 & \lambda^4 & \lambda^2 \\
      \lambda^4 & \lambda^2 & 1 \end{array} \right ) , \quad
  Y^d\sim  \lambda^2 \left ( \begin{array}{ccc}
      \lambda^6 & \lambda^4 & \lambda^4 \\
      \lambda^4 & \lambda^2 & \lambda^2 \\
      \lambda^4 & \lambda^2 & 1 \end{array} \right ),
\label{ypatterns}
 \ee
after breaking of $U(1)_A$ with $<\chi>/\bar M_P \sim<\chi'>/M_P \sim \
\lambda^2$. As we showed in Ref. \cite{js}, these patterns can fit the
data on quark masses and mixing.

We now wish to find appropriate leptonic charge assignments so that
$Y^L_{33}\sim Y^d_{33}$ and $\det M_d\sim \det M_L$. This model will have
$q_{_H}=1$ and an appropriate scale for $\mu_{eff}$.

Requiring $Y^L_{33}\sim Y^d_{33}$ means $\delta_\tau =-1$ or
$\delta_\tau=-3/2$.
Here we look only at the case $\delta_\tau=-1$. Using eqs.
(\ref{csigma}) and (\ref{qhsigmarel}), we deduce that
\be
a_1+a_2=\alpha_1+\alpha_2-{q_{_H}\over 6} = -{4\over 3}.
\label{a1a2rel}
 \ee
It is straightforward to classify all models which have $q_{_H}=1$ and
the above U(1)$_A$ charge assignments.  There are many such models
corresponding to different values of $a_1$ and $a_2$; however, not all of these
yield acceptable leptonic hierarchies.  One model which
gives an acceptable leptonic hierarchy has
\be
a_1=-{7\over 6},\quad a_2=-{1\over 6},
\label{achoices}
 \ee
so that
\be
 q(H_1 L_i e^c_j) = \left ( \begin{array}{ccc}
      -6 & -5  & -3.5 \\
      -5 & -4 & -2.5 \\
      -3.5 & -2.5 & -1 \end{array} \right ).
\label{qmodel}
\ee
The allowed Yukawa couplings are then
\be  Y^L_{ij}(\chi,\chi') \sim \left ( \begin{array}{ccc}
      (\chi')^4 & (\chi')^2\chi^2 & \chi'\chi^2 \\
      (\chi')^2\chi^2 & (\chi')^2\chi & \chi'\chi \\
      \chi'\chi^2 & \chi'\chi & \chi \end{array} \right ),
\label{allowed}
 \ee
giving rise to effective Yukawa terms
\be  Y^L\sim \lambda^2 \left ( \begin{array}{ccc}
      \lambda^6 & \lambda^6 & \lambda^4 \\
      \lambda^6 & \lambda^4 & \lambda^2 \\
      \lambda^4 & \lambda^2 & 1 \end{array} \right ),
\label{yuklep}
 \ee

There is no superpotential term of the form $H_1 H_2 \chi^n(\chi')^m,$ for
any $n,m\geq 0$ but there is a K\"ahler term
\be K\ni a {\chi^{\dag}\over \bar M_P } H_1 H_2 + h.c. ,\ee
which leads to
\be \mu_{eff}\sim a\lambda^2 m_{3/2}. \ee

 Once the previous charge assignments are given, eqs. (\ref{qqrel}) are three
linear equations for the average generational $U(1)_A$ charges. These equations
must be compatible with the solutions (\ref{sols}).
For example, for the second set of solutions
in (\ref{sols}) we find
\be
   x=-{29\over 15},\quad y=-{6\over 5}, \quad z={8\over 15},
\ee
for the explicit model given. These charge assignments have the feature that
the lepton and baryon number--violating superfield terms
\be \eta^i L_i H_2,\quad \lambda^{ijk}u^c_i d^c_j d^c_k, \quad
   \lambda'^{ijk}L_i Q_j d^c_k, \quad \lambda''^{ijk}L_i L_j e^c_k,
\label{blviolating}
 \ee
(where the coefficients are functions of $\chi,\chi'$)
are forbidden in perturbation theory, since one cannot render these terms
invariant under U(1)$_A$ using integral powers of fields.
For example, $q(L_1 H_2)=-5/6$, $q(L_2 H_2)=1/6$, $q(L_3 H_2)=5/3$,
so there is no integer power of $\chi$ and/or $\chi'$ which could couple to
these in a U(1)$_A$--invariant manner.

   In conclusion, we have discussed a class of models of fermion masses and
mixing based on a string-motivated U(1)$_A$ gauge symmetry
in which there is a natural
solution to the $\mu$ problem.  This solution implies a profound connection
between the fact that $\mu \sim v_{_{EW}}$ and the observed pattern of fermion
masses.

This research was partially supported by the NSF Grant PHY-93-09888.

\end{document}